\documentclass[useAMS,usenatbib]{mn2e}

\usepackage{times}
\usepackage{graphicx}

\newcommand{\beq}{\begin{equation}}
\newcommand{\eeq}{\end{equation}}

\def\gs{\mathrel{\lower0.6ex\hbox{$\buildrel {\textstyle >}\over{\scriptstyle \sim}$}}}
\def\ls{\mathrel{\lower0.6ex\hbox{$\buildrel {\textstyle <}\over{\scriptstyle \sim}$}}}

\begin{document}

\title[Dark matter vs MOND]{Dark matter vs. modifications of the gravitational inverse-square law. Results from planetary motion in the solar system}
\author[M. Sereno \& Ph. Jetzer]{M. Sereno$^{1}$\thanks{E-mail:
sereno@physik.unizh.ch (MS); jetzer@physik.unizh.ch (PhJ)} and Ph. Jetzer$^{1}$
\\
$^{1}$ Institut f\"{u}r Theoretische Physik, Universit\"{a}t Z\"{u}rich,
Winterthurerstrasse 190, CH-8057 Z\"{u}rich , Switzerland
}

\date{February 24, 2006}

\maketitle

\begin{abstract}
Dark matter or modifications of the Newtonian inverse-square law in the solar system are studied with accurate planetary astrometric data. From extra-perihelion precession and possible changes in the third Kepler's law, we get an upper limit on the local dark matter density, $\rho_\mathrm{DM} \ls 3 {\times} 10^{-16}~\mathrm{kg/m}^3$ at the 2-$\sigma$ confidence level. Variations in the $1/r^2$ behavior are considered in the form of either a possible Yukawa-like interaction or a modification of gravity of MOND type. Up to scales of $10^{11}$~m, scale-dependent deviations in the gravitational acceleration are really small. We examined the MOND interpolating function $\mu$ in the regime of strong gravity. Gradually varying $\mu$ suggested by fits of rotation curves are excluded, whereas the standard form $\mu(x) = x/(1+x^2)^{1/2}$ is still compatible with data. In combination with constraints from galactic rotation curves and theoretical considerations on the external field effect, the absence of any significant deviation from inverse square attraction in the solar system makes the range of acceptable interpolating functions significantly narrow. Future radio ranging observations of outer planets with an accuracy of few tenths of a meter could either give positive evidence of dark matter or disprove modifications of gravity.
\end{abstract}


\section{Introduction}

Gravitational inverse-square law and its relativistic generalization have passed significant tests on very different length- and time-scales. Precision tests from laboratory and from measurements in the solar system and binary pulsars provide a quite impressive body of evidence, considering the extrapolation from the empirical basis \citep{ade+al03,wil06}. First incongruences seem to show up only on galactic scales with the observed discrepancy between the Newtonian dynamical mass and the directly observable luminous mass and they are still in order for even larger gravitational systems. Two obvious explanations have been proposed: either large quantities of unseen `dark' matter (DM) dominate the dynamics of large systems \citep{zwi33} or gravity is not described by Newtonian theory on every scale \citep{fin63}. Dark matter is all the general theory of relativity needs to overcome apparent shortcomings and provides a coherent picture for gravitational phenomena from the laboratory to the cosmological context. The paradigm of cold DM when complemented with a positive cosmological constant (the so called $\Lambda$CDM scenario) is successful in explaining the whole range of galactic and extra-galactic body of evidence, from flat rotation curves in spiral galaxies to large scale structure formation and evolution \citep{pea99}. 

The $\Lambda$CDM paradigm could be regarded as the definitive picture apart from that the presumed existence of DM relies so long only on its putative global gravitational effect, whereas direct detection by any independent mean is still lacking. This makes room to alternative proposals based on modifications of Newtonian gravity. In general, such proposals do not extend the inverse-square law to a regime in which it has never before been tested and they do not introduce any exotic component. Proposals are very different from each other. Some of them can make gravity stronger on scales of galaxies and explain flat rotation curves without dark matter \citep{mil83}; others realize a mechanism for the cosmic acceleration without dark energy, for example as a result of gravity leaking on scales comparable to the horizon \citep{dva+al00}. Two main alternative proposals have been discussed. In the first one, the gravitational potential deviates from the usual form at large distances. A classical example is the inclusion of a Yukawa-like term in the gravitational potential. This is strictly related to more fundamental theories where these additional contributions appear as the static limit of interactions due to the exchange of virtual massive bosons \citep{ade+al03}. According to the second main choice, Newton's law fails when the gravitational acceleration is small rather than when the distance is large. The prototype and still one of the most empirically successful alternative to DM is Milgrom's modified Newtonian dynamics (MOND) \citep{mil83,sa+mc02}. With some basis in sensible physics, MOND can provide an efficient description of the phenomenology on scales ranging from dwarf spheroidal galaxies to cluster of galaxies but its cosmological extension is still in the childhood \citep{bek04}.

High precision solar system tests could provide model independent constraints on possible modifications of Newtonian gravity. The solar system is the larger one with very well known mass distribution and can offer tight confirmations of Newtonian gravity and general relativity. Any deviation emerging from classical tests would give unique information either on dark matter and its supposed existence or on the nature of the deviation from the inverse-square law. Several authors have discussed this possibility. \cite{tal+al88} derived limits from the analysis of various planetary astrometric data set on the variation in the $1/r^2$ behavior of gravity. Experimental bounds on non luminous matter in solar orbit were derived either by considering the third Kepler's law \citep{and+al89,and+al95} or by studying its effect upon perihelion precession \citep{gr+so96}. The influence of a tidal field due to Galactic dark matter on the motion of the planets and satellites in the solar system was further investigated by \cite{bra+al92} and \cite{kl+so93}. The orbital motion of solar-system planets has been determined with higher and higher accuracy \citep{pit05b} and recent data allow to put interesting limits on very subtle effects, such as that of a non null cosmological constant \citep{je+se06,se+je06}. In this paper, we discuss what state-of-art ephemerides tell us about non Newtonian or DM features. In section~\ref{basi}, we review standard expectations about Galactic dark matter at the solar circle and discuss some standard frameworks for deviations from the inverse square-law, i.e. a Yukawa-like fifth force and the MOND formalism. Observational constraints from perihelion precessions and changes in the third Keplerian law are discussed in section \ref{peri} and \ref{mean}, respectively. Section \ref{conc} is devoted to some final considerations.

\section{Basics}
\label{basi}

Let us now briefly consider the main features of dark matter in the solar system and of some alternatives to Newtonian gravity.

\subsection{Dark matter}

In the dark matter scenario, Milky Way is supposed to be embedded in a massive dark halo. Realistic models of the Milky Way based on adiabatic compression of cold DM haloes can be built in agreement with a full range of observational constraints \citep{kly+al02,ca+se05}. The local DM density at the solar circle is then expected to be $\rho_\mathrm{DM} \sim 0.2 \times 10^{-21}~\mathrm{kg/m}^3$, in excess of nearly five orders of magnitude with respect to the mean cosmological dark matter density.

\subsection{MOND}

MOND underpins the principle that gravitation departs from Newtonian theory if dynamical accelerations are small. It was initially proposed either as a modification of inertia or of gravity \citep{mil83}. According to this second approach, the gravitational acceleration $\mathbf{g}$ if related to the Newtonian gravitational acceleration $\mathbf{g}_\mathrm{N}$ as
\beq
\label{mond1}
 \mu (|\mathbf{g}|/a_0)\mathbf{g} = \mathbf{g}_\mathrm{N}
\eeq
where $a_0$ is a physical parameter with units of acceleration and $\mu (x)$ is an unspecified function which runs from $\mu (x)=x $ at $x \ll 1$ to $\mu (x) = 1 $ at $x \gg 1$. Whereas the Newtonian trend is recovered at large accelerations, in the low acceleration regime the effective gravitational acceleration becomes $g \simeq \sqrt{g_\mathrm{N} a_o}$. The asymptotically flat rotation curves of spiral galaxies and the Tully-Fisher law are explained by such a modification and a wide range of observations is fitted with the same value of $a_0 \simeq 1.2 \times 10^{-10}~\mathrm{m~s}^{-2}$ \citep{sa+mc02}. 

The $\mu$ function is formally free but, as a matter of facts, fits to rotation curves or considerations on the external field effects suggest a fairly narrow range \citep{zh+fa06}. The standard interpolating function proposed by \citet{mil83},
\beq
\label{mond2}
\mu (x) = x/\sqrt{1+x^2} ,
\eeq
provides a reasonable fit to rotation curves of a wide range of galaxies. Based on a detailed study of the velocity curves of the Milky Way and galaxy NGC~3198, \citet{fa+bi05} found out that interpolating functions which trigger a slower transition from the MONDian to the Newtonian regime should be preferred. They proposed the alternative interpolating function,
\beq
\label{mond3}
\mu (x) = x/(1+x) .
\eeq
Transition between the asymptotic regimes is smoother in Eq.~(\ref{mond3}) than in Eq.~(\ref{mond2}). In principle, $\mu$ could be precisely determined from the observations of an ideal galaxy in which both the flat rotation curve and the luminosity distribution are known with high accuracy. The $\mu$ function that best reproduces the Milky Way's rotation curve rotation seems to go smoothly from Eq.~(\ref{mond2}) at $x \ls 1$ to Eq.~(\ref{mond3}) at $x \gs 10$ \citep{fa+bi05}.

In the Newtonian regime, departures strongly depend on the way $\mu$ approaches 1 asymptotically. For a quite general class of interpolating functions, we can write \citep{mil83}
\beq
\label{mond4}
\mu (x) \simeq 1-k_0 (1/x)^m ,
\eeq
which leads to the modified gravitational field \citep{tal+al88}
\beq
\label{mond5}
\mathbf{g} \simeq \mathbf{g}_\mathrm{N}\left[ 1+ k_0 (a_0/ |\mathbf{g}_\mathrm{N}|)^m \right] .
\eeq
For $x \gg 1$, Eq.~(\ref{mond2}) and Eq.~(\ref{mond3}) can be recovered for $\left\{ k_0, m\right\}=\{1/2,2 \}$ and $\{1 ,1\}$, respectively.

Any viable relativistic theory embodying the MOND paradigm, such as Bekentein's TeVeS \citep{bek04} or Sanders'  BSTV \citep{san05}, seems to require scalar and vectorial fields in addition to the usual tensor field. MOND phenomenology emerges as an effective fifth force associated with a scalar field. The interpolating function $\mu$ is related to an auxiliary function of the scalar field strength. The parameterized post-Newtonian (PPN) formalism has been very effective in confronting metric theories of gravity with the results of solar-system experiments \citep{wil06}. Unfortunately, for the relativistic generalizations of MOND, the presence of both a scalar and a vector field, together with the free function in the Lagrangian that yields the expected dynamics in the low-acceleration limit, makes it problematic to derive the corresponding PPN parameters. To date, preliminary derivations only concern the very inner solar system, where $\mu$ is very close to unity \citep{bek04}, so that the very accurate determination of PPN parameters can not be directly used to test MOND.

\subsection{Yukawa-like fifth force}

Many long-range deviations can be characterized by an amplitude and a length scale. Let us consider additional contributions to the gravitational potential in the form of a Yukawa-like term, whose astrophysical consequences have been explored from the scale of the solar system \citep{ade+al03} to the large scale structure of the universe \citep{wh+ko01,am+qu04,sea+al05,shi+al05,se+pe06}. The weak field limit of the gravitational potential, $\phi$, can be written as a sum of a Newtonian and a Yukawa-like potential; for a point mass $M$,
\beq
\label{ykw1}
\phi  = -\frac{G_\infty M}{r}\left[ 1+ \alpha_\mathrm{Y}  \exp \left\{ -\frac{r}{\lambda_\mathrm{Y}}\right\}\right],
\eeq
where $\alpha_\mathrm{Y}$ is a dimensionless strength parameter and $\lambda_\mathrm{Y}$ is a length cutoff. The potential in Eq.~(\ref{ykw1}) goes as $\propto 1/r$ both on a small scale ($ r \ll \lambda_\mathrm{Y}$), with an effective coupling constant $G_\infty (1 +\alpha_\mathrm{Y})$, and on a very large scale, where the effective gravitational constant is $G_\infty $. We will take $G_\infty = G_\mathrm{N}/(1+\alpha_\mathrm{Y})$, so that the value of the coupling constant on a very small scale matches the observed laboratory value, $G_\mathrm{N}$. The total gravitational acceleration felt by a planet embedded in the potential~(\ref{ykw1}) can be written as,
\beq
\mathbf{g} = -\hat{\mathbf{r}}\frac{G_\infty M}{r^2}\left[ 1+\alpha_\mathrm{Y} \left(1+\frac{r}{\lambda_\mathrm{Y}} \right)  \exp \left\{ -\frac{r}{\lambda_\mathrm{Y}}\right\}  \right] .
\eeq
For $\alpha_\mathrm{Y}<0(>0)$, gravity is enhanced (suppressed) on a large scale. The potential in Eq.~(\ref{ykw1}) can be derived in a relativistic gravity model that obeys the equivalence principle \citep{zh+ne94}. A Yukawa-like contribution to the potential can be also connected to very-specific mass terms which appear in addition to the field theoretical analog of the usual Hilbert-Einstein Lagrangian \citep{ba+gr03}.

\section{Perihelion precession}
\label{peri}

\begin{table*}
\centering
\caption{\label{tab:peri} 2-$\sigma$ constraints from extra-precession of the inner planets of the solar system. $\delta\dot{\omega}_\mathrm{p}$ is the observed extra-precession rate from Pitjeva (2005b); $\delta {\cal A}_\mathrm{R}$  is a constant perturbative radial acceleration at the planet orbit and $\rho_\mathrm{DM}$ is the DM density within the planet orbit.}
\begin{tabular}{lrcl}
\hline
Name  &  $\delta\dot{\omega}_\mathrm{p}$~(arcsec/year) &
 $ \delta {\cal A}_\mathrm{R}~ (\mathrm{m}/\mathrm{s}^2) $  &  $ \rho_\mathrm{DM}~(\mathrm{kg}/\mathrm{m}^3)$ \\
\hline
Mercury  &  $-0.36(50)\times 10^{-4}$   &  $-1 {\times} 10^{-12} \ls \delta {\cal A}_\mathrm{R} \ls 5 {\times} 10^{-13}$   &   $ < 4 {\times} 10^{-14}$
\\
Venus    &  $0.53(30)\times 10^{-2}$    &  $-4 {\times} 10^{-12} \ls \delta {\cal A}_\mathrm{R} \ls 6 {\times} 10^{-11}$   & $ < 8 {\times} 10^{-14}$
\\
Earth    &  $-0.2(4)\times 10^{-5}$     &  $-5 {\times} 10^{-14} \ls \delta {\cal A}_\mathrm{R} \ls 3 {\times} 10^{-14}$  & $ <7{\times} 10^{-16}$
\\
Mars &      $0.1(5)\times 10^{-5}$      &  $-3 {\times} 10^{-14} \ls \delta {\cal A}_\mathrm{R} \ls 4 {\times} 10^{-14}$  &  $ < 3 {\times} 10^{-16}$
\\
\hline
\end{tabular}
\end{table*}

As well known, a test body moving under the influence of the Newtonian potential of a central mass $M$ will describe an ellipse with constant orbital elements. Due to a small, entirely radial perturbation, the argument of pericentre $\omega_\mathrm{p}$ will precess according to \citep[see][chapter~4]{sof89}
\beq
\label{peri1}
\dot{\omega}_\mathrm{p}=- \frac{(1-e^2)^{1/2}}{n a e}\delta {\cal A}_\mathrm{R} \cos f,
\eeq
where  $n \equiv \sqrt{G_\mathrm{N} M/a^3}$ is the mean motion of the unperturbed orbit, $a$ the semimajor axis, $e$ the eccentricity, $f$ the true anomaly counted from the pericentre and $\delta {\cal A}_\mathrm{R}$ the radial component of the perturbing acceleration. The longitude of the ascending node is not affected. 

Data from space flights and modern astrometric methods made it possible to create very accurate planetary ephemerides and to precisely determine orbital elements of solar system planets. The latest EPM2004 ephemerides were based on more than 317000 position observations collected over 1913-2003 and including radiometric and optical astrometric observations of spacecraft, planets, and their satellites \citep{pit05b}. Such ephemerides were constructed by simultaneous numerical integration of the equations of motion in the post-Newtonian approximation accounting for subtle effects such as the influence of 301 large asteroids and of the ring of small asteroids, as well as the solar oblateness. Extra-corrections to the known general relativistic predictions can be interpreted in terms of new physics. Results are listed in Table~\ref{tab:peri}. We considered the 2-$\sigma$ upper bounds. When the additional non-Newtonian acceleration is parameterized as constant, the average precession rate is given by
\beq
\label{peri2}
\langle \dot{\omega}_\mathrm{p} \rangle = \frac{(1-e^2)^{1/2}}{n a } \delta {\cal A}_\mathrm{R} .
\eeq
Best constraints on $\delta {\cal A}_\mathrm{R}$ come from Earth and Mars observations, see Table~\ref{tab:peri}. 

Analyzed data from Pioneer spacecrafts cover an heliocentric distance out to $\sim 70~\mathrm{AU}$ and show an anomalous acceleration directed towards the Sun with a magnitude of $\sim 9 {\times} 10^{-10}\mathrm{m~s}^{-2}$ which first appeared at a distance of 20~AU from the Sun \citep{and+al02}. If such an acceleration were gravitational in origin it would be not universal. In fact, effects on orbits of inner and outer planets would be large enough to have been detected given the present levels of accuracy \citep{and+al02,ior06,san06}. The upper bound from Mars in Table~\ref{tab:peri} is more than four orders of magnitude smaller then the Pioneer acceleration.

\subsection{Dark matter}

Galactic dark matter can cause extra-perihelion precession in the solar system. A refined analysis should consider the anisotropy in the gravitational field in the solar system due to tidal forces induced by the DM distribution \citep{bra+al92,kl+so93}. As an alternative approach, a spherically symmetric distribution around the Sun can be considered \citep{gr+so96,kh+pi06}. In fact, the effect at a given orbital radius is essentially given by the total DM mass contained within it, with a very weak dependence on the actual density distribution \citep{and+al89}. Dark matter density varies very slowly within the solar system and can be considered as nearly constant. Assuming a constant density  $\rho_\mathrm{DM}$, the perturbing radial acceleration at a radius $r$ is $\delta {\cal A}_\mathrm{R} = -(4\pi G_\mathrm{N} \rho_\mathrm{DM}/3) r $. After substituting in Eq.~(\ref{peri1}) and averaging over a period, the extra-precession rate can be written as
\beq
\label{peri3}
\langle \dot{\omega}_\mathrm{p} \rangle = - \frac{2 G_\mathrm{N} \pi \rho_\mathrm{DM}}{n}\left(1-e^2\right)^{1/2}.
\eeq
Note that for an effective uniform density of matter represented by a cosmological constant, i.e. $ \rho_\mathrm{DM} =-c^2 \Lambda/(4\pi G_\mathrm{N})$, the classical result for orbital precession due to $\Lambda$ is retrieved \citep{ker+al03,je+se06}. The best upper bound on local dark matter density comes from Mars data, see Table~\ref{tab:peri}. The accuracy on Mars precession should improve by more than six orders of magnitude to get constraints competitive with local estimates based on Galactic observables.

\subsection{MOND}

\begin{figure}
        \resizebox{\hsize}{!}{\includegraphics{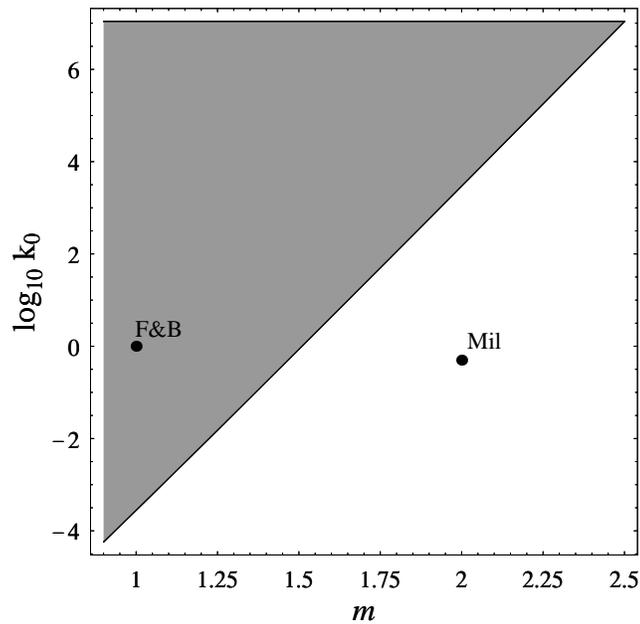}}
        \caption{Constraints on the MOND interpolating function, parameterized as in Eq.~(\ref{mond4}), arising from extra-perihelion precession of inner planets. The shadow region is ruled out at the  2-$\sigma$ confidence level. The points labeled Mil and F\&B locate the interpolating function in Eq.~(\ref{mond2}) and (\ref{mond3}), respectively.}
        \label{fig_peri_MOND_2sigma}
\end{figure}

The rate of perihelion shift in the Newtonian regime of MOND ($x \gg 1$) with a generic interpolating function in the form of Eq.~(\ref{mond4}) can be expressed in terms of hypergeometric functions. Here, we report only the case of a small eccentricity when
\begin{eqnarray}
\label{peri4}
\langle \dot{\omega}_\mathrm{p} \rangle & = & - k_0 n \left( \frac{a}{r_\mathrm{M}}\right )^{2 m} m \nonumber \\ 
& \times & \left\{ 1+ e^2 [ 1-m(5-2m)]/4 + {\cal O}(e^4) \right\},
\end{eqnarray}
where $r_\mathrm{M} \equiv \sqrt{G_\mathrm{N} M/a_0}$. As for the DM case, the Mars data is the more effective in constraining the parameter space, see Fig.~\ref{fig_peri_MOND_2sigma}. For $k_0 \sim 1$, we get $m \gs 1.5$. Results from solar system are in disagreement with expectations based on the extrapolation to the strong acceleration regime of the free functions preferred on a galactic dynamics basis. From the study of rotation curves, accelerations seems to continue to increase quite smoothly as $1/r$ even in the intermediate MONDian regime. As a consequence, for  $ x \ls 1$ free functions which trigger a smooth transition, as Eq.~(\ref{mond3}), must be preferred over expressions such as Eq.~(\ref{mond2}) \citep{fa+bi05,zh+fa06}. On the other hand, in the Newtonian regime, $x \gg 1$, the functional form of Eq.~(\ref{mond3}) is clearly excluded by solar system data, whereas Eq.~(\ref{mond2}) is still compatible. Combining data from solar system and galactic dynamics, in the comparison between  Eq.~(\ref{mond2}) and  Eq.~(\ref{mond3}), the first one seems to be preferred in both the deep MONDian and Newtonian regimes, whereas the second one gives a better fit for the intermediate region. Similar considerations induced \cite{san06} to argue that the total gravitational acceleration is strictly Newtonian, i.e. $\propto 1/r^2$, on small scales and that the transition to the total asymptotic acceleration $\propto 1/r$ shows up through a plateau region between $10^2$ and $10^3$~AU where the extra acceleration is more or less constant. The accuracy on Mars data should improve by nearly four orders of magnitude to disprove the standard interpolating function in Eq.~(\ref{mond2}).

\subsection{Yukawa fifth force}

\begin{figure}
        \resizebox{\hsize}{!}{\includegraphics{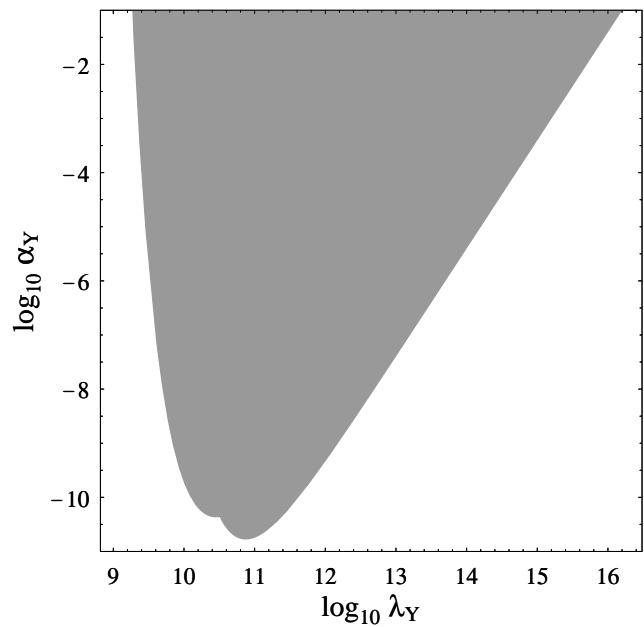}}
        \caption{Constraints on the Yukawa-like fifth force parameters, for positive $\alpha_\mathrm{Y}$, arising from extra-perihelion precession of inner planets. The shadow region is ruled out at the  2-$\sigma$ confidence level.}
        \label{fig_peri_yuk_pos_2sigma}
\end{figure}

\begin{figure}
        \resizebox{\hsize}{!}{\includegraphics{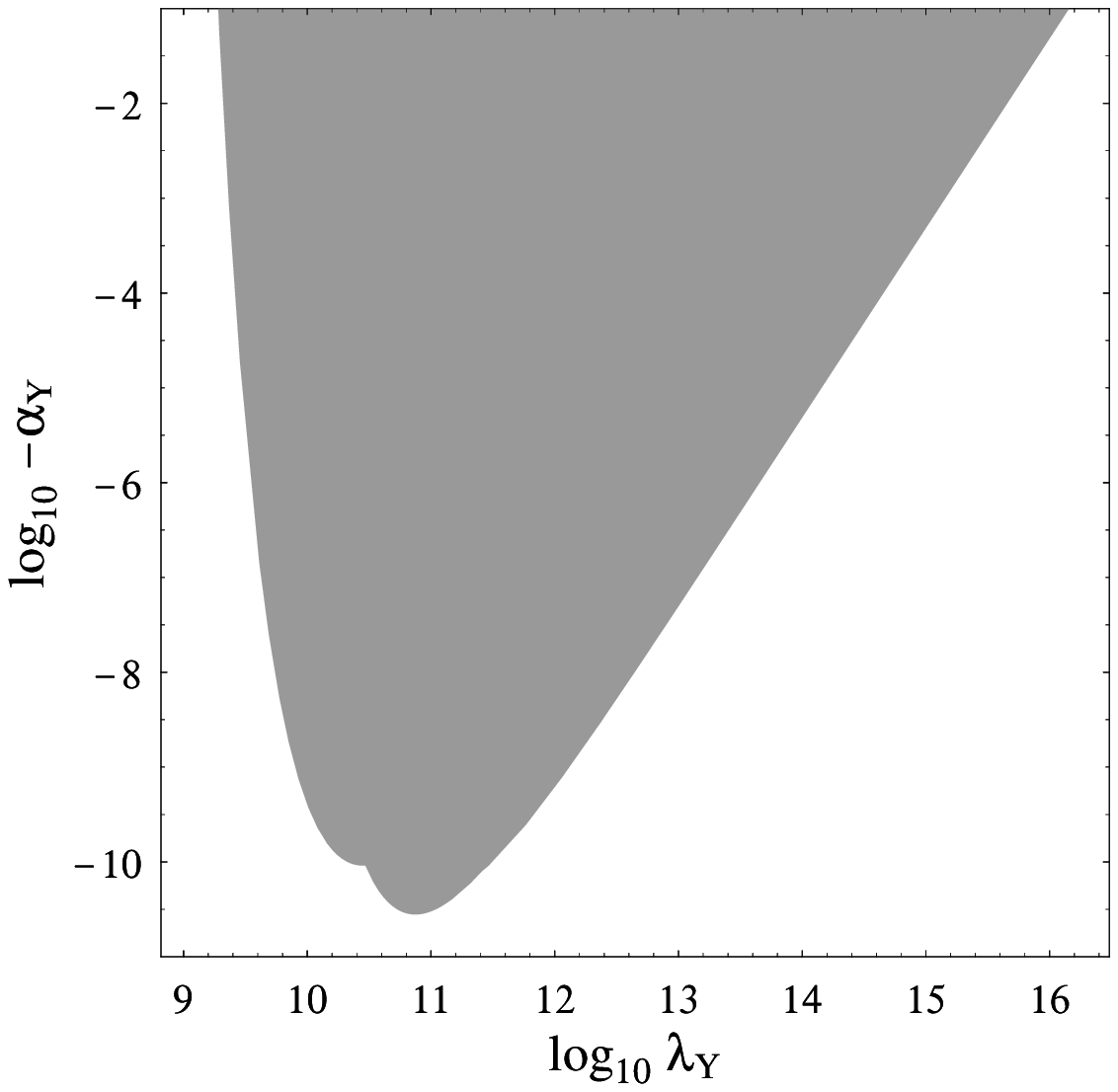}}
        \caption{Constraints on the Yukawa-like fifth force parameters, for negative $\alpha_\mathrm{Y}$, arising from extra-perihelion precession of inner planets. The shadow region is ruled out at the  2-$\sigma$ confidence level.}
        \label{fig_peri_yuk_neg_2sigma}
\end{figure}

The anomalous precession rate due to a Yukawa-like contribution to the gravitational potential is
\begin{eqnarray}
\label{peri5}
\langle \dot{\omega}_\mathrm{p} \rangle & = & \alpha_\mathrm{Y} \left( \frac{a}{\lambda_\mathrm{Y}} \right)^2 \exp\left\{ -\frac{a}{\lambda_\mathrm{Y}}\right\}\frac{n}{2} \\
& \times & \left\{ 1-\frac{1}{8} \left[ 4 - \left(\frac{a}{\lambda} \right)^2 \right]e^2  + {\cal O} (e^4)\right\}. \nonumber
\end{eqnarray}
Extra-precession data for a planet with semimajor axis $a$ mainly probe scale lengths of $\lambda_\mathrm{Y} \sim a/2$. Solar system data allow to constrain departures from the inverse-square law with high accuracy for a scale length $\lambda_\mathrm{Y} \sim 10^{10}-10^{11}$~m \citep{tal+al88,ior05}. Bounds are mainly determined from Mercury and Earth data, see Figs.~\ref{fig_peri_yuk_pos_2sigma} and~\ref{fig_peri_yuk_neg_2sigma}. For $\lambda_\mathrm{Y} \sim 10^{11}$~m, we get $-5 \times 10^{-11} \ls \alpha_\mathrm{Y} \ls 6 \times 10^{-11}$.

\section{Third Kepler's law}
\label{mean}

\begin{table}
\caption{\label{tab:mean} 2-$\sigma$ upper bounds from anomalous mean motion of the solar system planets; $\delta a$ is the uncertainty on the semimajor axis from Pitjeva (2005a); $\delta {\cal A}_\mathrm{R}$ is an anomalous constant radial acceleration; $\rho_\mathrm{DM}$ is the dark matter density.}
\begin{tabular}{lrrl}
\hline
Name  &  $ \delta a~(\mathrm{m})$  & $ |\delta {\cal A}_\mathrm{R} |~ (\mathrm{m}/\mathrm{s}^2) $  &  $ \rho_\mathrm{DM}~(\mathrm{kg}/\mathrm{m}^3)$ \\
\hline
Mercury  &  $0.105 \times 10^{+0}$   &  $ \ls 4 {\times} 10^{-13}$   &   $  \ls 3 {\times} 10^{-14}$ \\
Venus    &  $0.329 \times 10^{+0}$   &  $ \ls 2 {\times} 10^{-13}$   &   $  \ls 7 {\times} 10^{-15}$ \\
Earth    &  $0.146 \times 10^{+0}$   &  $ \ls 3 {\times} 10^{-14}$   &   $  \ls 8 {\times} 10^{-16}$  \\
Mars &      $0.657 \times 10^{+0}$   &  $ \ls 4 {\times} 10^{-14}$   &   $  \ls 7 {\times} 10^{-16}$ \\
Jupiter &   $0.639  \times 10^{+3}$  &  $ \ls 1 {\times} 10^{-12}$   &   $  \ls 5 {\times} 10^{-15}$\\
Saturn  &   $0.4222 \times 10^{+4}$  &  $ \ls 1 {\times} 10^{-12}$   &   $  \ls 3 {\times} 10^{-15}$\\
Uranus  &   $0.38484\times 10^{+5}$  &  $ \ls 1 {\times} 10^{-12}$   &   $  \ls 2 {\times} 10^{-15}$ \\
Neptune &   $0.478532\times 10^{+6}$ &  $ \ls 4 {\times} 10^{-12}$   &   $  \ls 3 {\times} 10^{-15}$  \\
Pluto   &  $0.3463309\times 10^{+7}$ &  $ \ls 1 {\times} 10^{-11}$   &   $  \ls 8 {\times} 10^{-15}$  \\
\hline
\end{tabular}
\end{table}

A departure from the  inverse-square law could affect the radial motion of a body around a central mass and a change in the Kepler's third law would occur. The Newtonian law of motion for a test body in a circular orbit around a central mass $M$, in presence of a perturbing radial acceleration $\delta {\cal A}_\mathrm{R}$, can be written as
\begin{eqnarray}
\omega^2 r & =&  \frac{G_\mathrm{N} M}{r^2} - \delta {\cal A}_\mathrm{r}  \label{mean1} \\
& \equiv & \frac{G_\mathrm{N} M_\mathrm{eff}}{r^2} . \nonumber
\end{eqnarray} where $\omega$ is the angular frequency and $ M_\mathrm{eff} \equiv M (1+\delta {\cal A}_\mathrm{r}/{\cal A}_\mathrm{N})$ is the effective mass felt by the orbiting planet. In other words, the angular frequency will differ from the mean motion $ n \equiv \sqrt{G_\mathrm{N} M/ a^3}$. It is
\begin{equation}
\label{mean4}
\frac{\delta n}{n} =  \frac{1}{2} \frac{\delta {\cal A}_\mathrm{r}}{{\cal A}_\mathrm{N}}.
\end{equation}
Variation of the effective solar mass felt by the solar system inner planets with respect to the
effective masses felt by outer planets could probe new physics \citep{and+al89,and+al95}. We can evaluate the statistical error on the mean motion for each major planet from the uncertainty on the semimajor axis, $\delta n = - (3/2) n \delta a/a$, and translate it into an uncertainty on the effective acceleration. Results for a constant additional acceleration term are listed in Table~\ref{tab:mean}. Assuming $\delta {\cal A}_\mathrm{r}$ being constant, it would be $\ls 5 \times 10^{-12}\mathrm{m}/\mathrm{s}^2$ in the range 20-30~AU, as can be inferred from Uranus and Neptune orbits, well below the anomalous Pioneer acceleration \citep{ior06}.
Limits from Earth and Mars are competitive with data from perihelion precession. Errors in Table~\ref{tab:mean} are formal and could be underestimated. Current accuracy can be determined evaluating the discrepancies in different ephemerides. Differences in the heliocentric distances do not exceed 10~km for Jupiter and amount to 180, 410, 1200 and 14000~km for Saturn, Uranus, Neptune and Pluto, respectively \citep{pit05a}. Bounds from outer planets reported in Table~\ref{tab:mean} should be accordingly increased.

\subsection{Dark matter}

Bounds on $\rho_\mathrm{DM}$ from deviations in the mean motion of inner planets, see Table~\ref{tab:mean}, are of the same order of magnitude of constraints from extra-precession. Observations of outer planets provide constraints that are an order of magnitude larger but they give the best future prospects. Unlike inner planets, radio-technical observations of outer planet are still missing and their orbits can not be determined with great accuracy. Since the required accuracy to probe the effects of a given uniform background decreases as $\propto a^{-4}$, whereas the measurements precision of ranging observations is roughly proportional to the range distance, exploration of outer planets seems pretty interesting. Dark matter with $\rho_\mathrm{DM} \simeq 0.2 \times 10^{-21}~\mathrm{kg/m}^3$ could be detected if the orbital axis of the Uranus, Neptune and Pluto orbits were determined with an accuracy of $\delta a \sim 3 \times 10^{-2},~ 2\times 10^{-1}$ and $5\times 10^{-1}$~m, respectively. Up till now, the only ranging measurements available for Uranus and Neptune are the Voyager 2 flyby data, with an accuracy in the determination of distance of $\sim 1~\mathrm{km}$ \citep{and+al95}, not so far from what required to probe solar system effects of dark matter.\footnote{The approach followed here is more conservative than a similar analysis appeared in \cite{ior06b}, where it is assumed that all of the observational residuals can be fully accounted for with a suitable combination of different effects. Furthermore, \cite{ior06b} considers a quite peculiar form of the gravitational potential valid only for a dark mass distribution fully contained within the considered orbit.}

\subsection{MOND}

Results from analysis of mean motion are similar to extra-precession analysis. The interpolating function in Eq.~(\ref{mond3}) is not consistent with solar system data. From Uranus data, we get  $m \gs 1.4$ assuming $k_0 \simeq 1$. Again, best future prospects are related to radio-technical determination of orbits of outer planets. The standard interpolating function in Eq.~(\ref{mond2}) could be (dis-)probed if the axes of the Uranus, Neptune and Pluto orbit were determined with an accuracy of $\delta a \sim 3 \times 10,~ 3\times 10^{2}$ and $ 1\times 10^{3}$~m, respectively.

\subsection{Yukawa fifth force}

\begin{figure}
        \resizebox{\hsize}{!}{\includegraphics{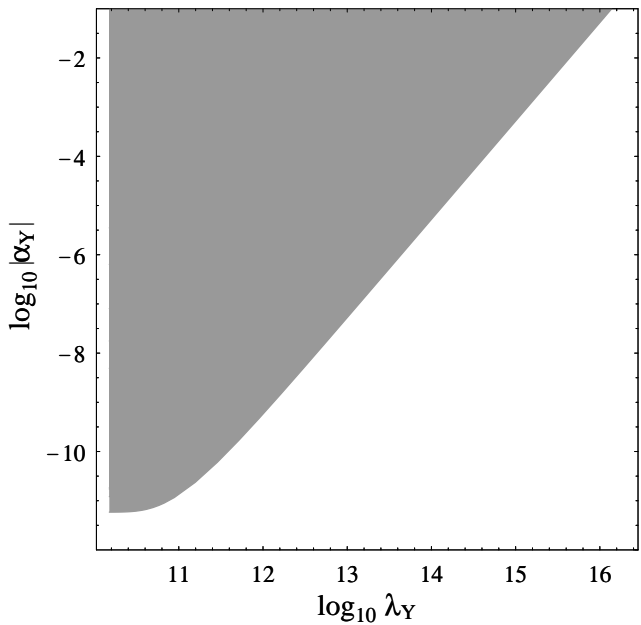}}
        \caption{Constraints on Yukawa fifth force parameters, for the absolute value of $\alpha_\mathrm{Y}$, arising from deviations from the third Kepler's law. The shadow region is ruled out at the 2-$\sigma$ confidence level.}
        \label{fig_mean_yuk_pos_2sigma}
\end{figure}

Comparison of Keplerian mean motions of inner and outer planets can probe a Yukawa-like contribution only if planets feel different effective gravitational constants. Such test is insensitive to values of $\lambda_\mathrm{Y}$ either much less the orbit radius of the inner planet or much larger than the orbit of the outer planets \citep{ade+al03}. Differently from extra-precession of perihelion, which appears only for departures from the inverse square law, changes in the mean motion can appear even if both planets feel a gravitational acceleration $\propto 1/r^2$  but with different renormalized gravitational constants. Considering inner planets, Earth data give $|\alpha_\mathrm{Y}| \ls 6 \times 10^{-12}$ for $\lambda_\mathrm{Y} \ls 2 \times 10^{10}$~m. The best constraint from outer planets is due to Jupiter, with $|\alpha_\mathrm{Y}| \ls 5 \times 10^{-9}$ for $\lambda_\mathrm{Y} \ls 10^{11}$~m.

\section{Conclusions}
\label{conc}

Debate between dark matter and departures from inverse-square law is still open. Considering both theoretical and observational aspects, dark matter seems to be slightly preferred. If on a galactic scale the two hypotheses match, on the cosmological side only DM can give a consistent framework. This might shortly change with the steady improvements in relativistic generalization of the MONDian paradigm. So, in our opinion, it is of interest to examine results on a very different scale, that of the solar system. Solar system data have been confirming predictions from the general theory of relativity without any need for dark matter and it is usually assumed that deviations can show up only on a larger scale. In this paper, we have explored what we can learn from orbital motion of major planets in the solar system. Results are still non-conclusive but nevertheless interesting. Best constraints come from perihelion precession of Earth and Mars, with similar results from modifications of the third Kepler's law. The upper bound on the local dark matter density, $\rho_\mathrm{DM} \ls 3 {\times} 10^{-16}~\mathrm{kg/m}^3$, falls short to estimates from Galactic dynamics by six orders of magnitude.  

Deviations of the gravitational acceleration from $1/r^2$ are really negligible in the inner regions. A Yukawa-like fifth force is strongly constrained on the scale of $\sim 1$~AU. For a scale-length $\lambda_\mathrm{Y} \sim 10^{11}~\mathrm{m}$, a Yukawa-like modification can contribute to the total gravitational action for less then one part on $10^{11}$. Similar limits could be achieved by precise measurements on the proof masses carried on board of the LISA Pathfinder satellite~\footnote{http://www.rssd.esa.int/index.php?project=LISAPATHFINDER} (Speake, private communication). In fact, instantaneous measurements of the drag-free test-mass acceleration during the transfer orbit towards the first Sun-Earth Lagrange point could in principle test the inverse square law on a scale length of $\sim 1$~AU (Speake, private communication). Results on a similar scale-length could be obtained through a detailed analysis of binary pulsars. The periastron shift, the gravitational redsfhift/second-order Doppler shift parameter and the rate of change of orbital period are sensitive to scalar-tensor gravity and to any other deviation from the general theory of relativity \citep{wil06}. Dipole gravitational radiation associated with violations of the equivalence principle in its strong version could cause an additional form of gravitational damping and a significant change of the orbital period could occur, in particular for a binary pulsar system with objects of very dissimilar mass \citep{wil06}. A massive graviton associated with a Yukawa-like fifth force could also affect the speed of propagation of gravitational waves and induce radiation effects at the reach of future gravitational wave detectors \citep{wil06}.

A large class of MOND interpolating function is excluded by data in the regime of strong gravity. The onset of the asymptotic $1/r$ acceleration should occur quite sharply at the edge of the solar system, excluding the more gradually varying $\mu(x)$ suggested by fits of rotation curves. On the other hand, the standard MOND interpolating function $\mu(x) = x/(1+x^2)^{1/2}$ is still in place. Studies on planetary orbits could be complemented with independent observations in the solar system. Mild or even strong MOND behavior might become evident near saddle points of the total gravitational potential, where MONDian phenomena might be put at the reach of measurements by spacecraft equipped with sensitive accelerometers \citep{be+ma06}. As a matter of fact, fits to galactic rotation curves, theoretical considerations on the external field effects and solar system data could determine the shape of the interpolating function with a good accuracy on a pretty large intermediate range between the deep Newtonian and MONDian asymptotic behaviors.

Future experiments performing radio ranging observations of outer planets could greatly improve our knowledge about gravity in the regime of large accelerations. The presence of dark matter could be detected with a viable accuracy of few tenths of a meter on the measurements of the orbits of Neptune or Pluto, whereas an uncertainty as large as hundreds of meters would be enough to disprove some pretty popular MOND interpolating functions. 

In order to become really competitive with general relativity and the $\Lambda$CDM paradigm, MOND should be predictive on the whole range of observed systems from solar system to the cosmic microwave background radiation. On a galactic scale, effects of DM or MOND are pretty similar and very difficult to distinguish each other but there might be some detectable differences on a smaller scale. In fact, the local value of DM at the solar circle is pretty much fixed by Galactic dynamics whereas the MOND behavior in the regime of strong accelerations probed locally is not univocal on a theoretical and observational basis. Nevertheless, only a very small class of interpolating free functions would give the same perturbation on the orbits of outer planets as that from local DM. Matching the expectations from DM with future radio ranging observations would be an important, nearly conclusive confirmation of its existence. On the other hand, deviations at a different order of magnitude, as those expected for a large variety of MOND interpolating functions, would be a strong indication of departure from the inverse-square gravitational law.

\section*{Acknowledgments}
The authors thank Clive C. Speake for information on the LISA Pathfinder capabilities to test Yukawa-like forces and an anonymous referee for constructive criticism. M.S. is supported by the Swiss National Science Foundation and by the Tomalla Foundation.


\setlength{\bibhang}{2.0em}

\end{document}